\documentclass[preprintnumbers]{revtex4}
\usepackage{natbib}
\usepackage{amsmath}
\usepackage{amstext}
\usepackage{amssymb}
\usepackage{latexsym}
\usepackage{theorem}
\usepackage{placeins}
\usepackage{moreverb}
\usepackage[amssymb]{SIunits}
\usepackage{xspace}
\usepackage{epsfig}
\usepackage{wrapfig}
\usepackage{hyperref}
\usepackage{epstopdf}
\usepackage[none]{hyphenat}

\usepackage{color}
\usepackage{soul}

\newcommand{\ie}{i.e.{}\xspace}			
\newcommand{\cf}{cf.{}\xspace}			

\newcommand{\viceversa}{\emph{vice versa}\xspace}

\newcommand{\Exp}[1]{\langle #1 \rangle}

\newcommand\bra[1]{\ensuremath{\langle#1|}}
\newcommand\ket[1]{\ensuremath{|#1\rangle}}
\newcommand\iprod[2]{\ensuremath{\langle #1 | #2 \rangle}}

\newcommand{\LI}{\begin{itemize}}
\newcommand{\IE}{\end{itemize}}
\newcommand{\LN}{\begin{enumerate}}
\newcommand{\NE}{\end{enumerate}}
\newcommand{\LD}{\begin{description}}
\newcommand{\DE}{\end{description}}
\newcommand{\LL}{\begin{list}}
\newcommand{\LE}{\end{list}}

\newcommand\Section[2]{\section{#2}\label{s:#1}}

\begin{document}
\title{The equilibrium classical scatter spectrum of waves}
\author{V. Guruprasad}
\affiliation{Inspired Research, New York.}
\begin{abstract}
\noindent
Regardless of
	the unspecific notions of photons as
		light complexes, radiation bundles or wave packets,
the radiation from
	a single state transition
is at most
	a single continuous wave train
that starts and ends
	with the transition.
The radiation equilibrium spectrum must be
	the superposition sum of
		the spectra of such wave trains.
A classical equipartition of
	wave trains
cannot diverge
	since they would be finite in number,
whereas standing wave modes are
	by definition infinite,
		which had doomed Rayleigh's theory,
and
	concern only
		the total radiation.
Wave trains are
	the microscopic entities of radiation
		interacting with matter,
that correspond to
	molecules in kinetic theory.
Their quantization
	came from matter transitions
in Einstein's 1917 derivation
	of Planck's law.
The spectral scatter of wave trains
	by Doppler shifts,
which cause
	the wavelength displacements in Wien's law
used for the frequency dependence
	in Einstein's derivation,
is shown to yield
	the shape of the Planck spectrum.
A Lorentz transform property of Doppler shifts
	discovered by Einstein
is further shown equivalently
	necessary and sufficient
		to have corrected Rayleigh's theory.

\end{abstract}
\maketitle

\newcommand\Expv[1]{\ensuremath{\bar{#1}}}
\newcommand{\ex}{\mathbf{e}_x}
\newcommand{\ey}{\mathbf{e}_y}

\Section{intro}{Introduction} 

Presented below is
	the first-ever treatment of
the classical spectral scatter of
	electromagnetic wave trains
		in a cavity,
due to classical Doppler shifts by simple reflection
	from atoms or cavity walls
and
	their thermal motions or vibrations.
As the number and energy of
	wave trains
from
	a finite set of atoms
		would be finite,
the divergence of
	Rayleigh's theory
		cannot occur.
The spectral scatter is shown to bear
	the form of the Planck spectrum
classically due to
	proportional energy increases
as a Lorentz transform effect in
	Doppler frequency shifts,
incidentally discovered by Einstein
	\cite{Einstein1905c},
who ascribed it to
	``light complexes''
	\cite{Einstein1905c}
	or ``bundles''
	\cite{Einstein1917qt}
in deference to
	the quantum hypothesis
	\cite{Redzic2004,Norton2006,Norton2008}.
Other traditional notions of photons as wave packets having
	distinct phase and group velocities
	(\cf \cite[\S{31}]{Dirac})
		refer to Fourier spreads
	also possessed by
		the wave trains.
There was also a recent attempt to identify photons with
	the individual undulations in sinusoidal waves
	\cite{Popescu2007},
which makes partial sense in
	harmonic families of standing wave modes
	\cite{Prasad2006b}.
The present focus is more simply on
	\emph{arbitrary} classical travelling waves
constrained only by finiteness of
	length and energy.

The Lorentz property holds analytically for
	classical electromagnetic waves of
		arbitrary amplitudes
	between
		phase boundaries
	\cite{Popescu2007},
and thus specifically
	the wave trains.
Einstein's notions of
	radiation energy and momentum in
		detailed balance with state transitions in matter 
	\cite{Einstein1917qt}
extended
	known classical principles of these quantities
		to radiation,
and its only other input
	concerning radiation was Wien's law,
		also classical in origin.
His treatment was thus
	a rigorous classical derivation of
		Planck's law
that traced
	quantization to matter transitions,
despite
	his own apparent attribution of the Lorentz property
		to inherent quantumness.

The present result proves
	this contrarian insight correct
by showing that
	the same physical mechanisms
do reproduce
	the Planck form
		without quantization
in the absence of
	matter transitions.
It also implies that
	Einstein's treatment was incomplete,
despite reproducing
	Planck's law,
in overlooking
	the Doppler shifts from non-transitional reflections,
and that
	radiation quantization is
		an over-reaching premise
though empirically correct at 
	optical frequencies.

The result complements
	emerging reports of entanglement
obtainable classically
	via fluid mechanics
	\cite{Brady2013}
with relation to
	electromagnetic theory
	\cite{Brady2015},
and from Maxwell's equations
	under various conditions
	\cite{Carroll2015,Qian2013},
even as
	loophole-free confirmations of non-locality
		are achieved for the quantum result
	\cite{LoopholeFree2015}.
The quantumness implication of entanglement itself
	is now questioned
	\cite{Ghose2013,Qian2015},
echoing
	general anticipation of entanglement in classical waves
in
	\cite{Prasad2000b}.

The result is
	straightforward,
and could be overturned only by
	an error in assuming
that
	any single state transition of matter
emits at most
	a single continuous wave train of radiation.
If we assumed,
	like Planck and Einstein,
that matter comprises 
	a dynamical structure of particles,
		some with charge,
and its internal states comprise
	specific arrangements and motions,
then classically,
	a state transition must involve
		a continuous change
	from one arrangement and motions to the next,
implying
	a continuous variation of
		electric and magnetic forces
	at any point of observation,
which qualifies as
	a continuous wave train.
Emission from
	a single state transition
cannot be
	a multitude of whole wave trains,
nor
	a more general combination of
		sinusoidal wave components
	amounting to more than
		a single wave train,
as also inferred by Einstein
	\cite{Einstein1917qt}
	for concurrently acting ``bundles''.

A finite wave train is
	the simplest and most general description of
an electromagnetic waveform
	associable with
		any single, discrete state transition in matter,
and is sufficient, 
as each such wave train would not only
	bear a full energy quantum
but also satisfy
	directionality and momentum balance
		in interactions with matter,
	as shown by Einstein
	\cite{Einstein1917qt},
and its spatial extent suffices to explain
	quantum non-locality,
		as also proved here.
The traditional broader allowance for complexes or bundles
	thus did not serve for generality
except in analogy to
	``complexions'' or microstates
in view of
	their arbitrary
		amplitude and polarization distributions
	and
		wavefronts,
and has instead inadvertently denied
	the only possible functional form for representing
		discrete interactions with matter.
Related historical difficulties,
leading to a belief in
	the inadequacy of classical physics itself,
can be attributed to
	standing wave modes
		without the Lorentz property
being
	the only classical model
		in Rayleigh and Planck theories
	\cite{Planck1900,Planck1901,Planck1914},
as well as
in recent ideas of
	a classical origin of Planck's law arguing
		zero-point energy and Casimir forces are classical
	(\cf \cite{Boyer1969,Boyer2002,Boyer2003,Boyer2012,Cercignani1972,Cercignani1998}).

The modes are favoured as
	a stationary representation of
		the equilibrium steady state,
and also
	for their similarity to
		the vibrational modes of molecules
and to
	cyclic coordinates associated with stationary states
		in classical equipartition
	\cite{Rayleigh1900ke,Rayleigh1900rad}.
However,
	standing wave modes
by definition concern
	the \emph{total} radiation in the cavity,
and the role of molecular vibrations
	more precisely concerns
		spin and orbital angular momenta,
	as contained in
		the polarization and wavefront distributions,
	respectively
which are now well understood
	\cite{NaturePhotonics2015}.
Further,
	modes are defined by geometry
		and inherently infinite in number,
as
	wavelengths are infinitely divisible
		real valued quantities,
so Rayleigh's theory was
	doomed to divergence by
		this choice of entities,
in treating
	classical equilibration
		without the Lorentz property.
As shown later by Bose
	\cite{Bose1924},
Planck's oscillator hypothesis
	also obviates Wien's law,
and thereby
	\emph{all} dynamical considerations.
This leaves Einstein's work
	the only historical consideration of
		the microscopic dynamics of radiation.

Correspondingly,
the present theory is the radiative analogue of
	the treatment of molecular motions
		in the kinetic theory
but got skipped in
	radiation theory
due to
	the success of
		macroscopic thermodynamic principles,
	using standing waves,
in yielding
	the analogous result of
		radiation pressure
in
	Stefan-Boltzmann and Wien's laws.
The wavelength displacements
	in Wien's law
are Doppler shifts upon reflection at walls 
	subject to macroscopic motions
	\cite{Planck1914},
so wall Doppler shifts provided
	a critical physical mechanism for dispersing
		the radiant energy across frequencies
	in treatments
		including Einstein's,
as modes,
	by definition,
cannot mutually interact 
	and atomic spectra are narrow.
However,
as all considerations in Wien's law
	are macroscopic,
it constrained rather than explain
	the spectral equilibration.
Brownian motions
	due to radiation reactions,
also considered by Einstein
	\cite{Einstein1917qt},
would be
	a weak mechanism at very low pressures.

It could be argued,
	especially on the basis of Fermi-Pasta-Ulam theory
	(\cf \cite{FPU1955,TodaKuboSaito1992,Berman2005,Porter2009})
	and related ideas of fluctuations,
that reflections by stationary walls suffice
	to explain the spectral equilibration,
as they would spatially disperse
	wave fronts much like molecules.
However,
the premise of
	stationary confining walls
in both kinetic and radiation theories is itself
	another traditionally overlooked defect,
since wall impacts,
	whether by molecules or photons,
should also set
	real walls of finite mass
		into vibration,
much as
	the photon interactions would set
		the molecules into Brownian motions
	in Einstein's treatment
	\cite{Einstein1917qt},
until
	the wall vibration modes reach equilibrium with
		the confined gas or radiation.
The wall vibrations do not alter
	equilibrium energy, temperature or pressure,
so their neglect
	did not affect the overall result,
but their absence must denote
	walls at absolute zero temperature
		by the third law,
	which was only realized later
		\cite{Nernst1926},
and would thereby violate
	the premise of equilibrium of
		a confined gas or radiation
			at a steady non-zero temperature.
Their inclusion is thus required for
	completeness of the dynamical considerations,
and would also resolve
	Loschmidt's paradox
	(\cf \cite{Loschmidt1876,PhysWebPlanck2000})%
	\footnote { 
	Essentially by implying that
		a statistical system in equilibrium cannot be also truly closed.
	}, 
besides providing
	a general mechanism
for Doppler shifts
	at the microscopic level.

The overall result is thus not only that
	the classical electromagnetic wave
is also
	the most general notion of a photon
		given by quantum theory,
in all aspects of
	matter interactions and thermodynamics
as well as
	non-locality,
but more so that
	the current ideas fundamentally derive from
		ignorance and flaws
	in the classical attempts,
and misperception of
	a key classical electromagnetic wave property
		by Einstein.
Complementary ideas of
	the thermodynamics of
		information representation
	and
		erasure
	in physical observers
	\cite{Landauer1961b,Landauer1991,Bennett1973,Bennett1987},
and application of
	the third law to
		observer state transitions
	in process of physical observations,
classically impose
	an inherent probabilistic character
		to all observations,
so that
	even the quantum probabilities
cannot remain
	a matter of postulate.


\vspace{3pt}

The next section revisits
	Einstein's 1905 and 1917 papers
		in the present view.
Sufficiency of the Lorentz property
	to correct Rayleigh's theory
is shown
	in Section \ref{s:rayleigh}.
The core result for wave trains follows
	in Section \ref{s:theory}.


\Section{background}{Background of the core result} 

In the 1917 paper
	\cite{Einstein1917qt},
Einstein considered that
	the equilibrium with radiation
could not change
	the molecular energy distribution,
	$
	W_n
	=
		p_n
		e^{
			- E_n  k_B T
		}
	$
	from the kinetic theory,
where
	$W_n$ denoted the relative frequency of
		a state of energy $E_n$.
This led,
	via considerations of energy and momentum exchanges
		with radiation,
to the condition
\begin{equation} \label{e:einstein} 
		p_n
		e^{
			- E_n  k_B T
		}
		B^{m}_{n}
		\rho
	=
		p_m
		e^{
			- E_m  k_B T
		}
		\left(
			B^{n}_{m}
			\rho
		+
			A^{n}_{m}
		\right)
\quad
\text{whence}
\quad
	\rho
	=
		\frac{
			A^{n}_{m}  B^{n}_{m}
		}{
			e^{
				(
				E_m
				-
				E_n
				)
				 k_B T
			}
			- 1
		}
\end{equation} 
where
	$A^{n}_{m}$ was
		a probability coefficient to represent
			spontaneous emission,
and
	$B^{m}_{n}$ and $B^{n}_{m}$ were
		coefficients to denote
			absorption and stimulated emission,
	respectively.
The emission and absorption mechanisms
	were the only assumptions
		new in the 1917 paper,
and no inherent dependence on frequency
	was assumed.

The connection to frequencies was obtained by
equating to
	Wien's law in the form
	$
	\rho =
		\alpha \nu
		e^{
			- h \nu  k_B T
		}
	$,
which led to
	the relations
	$
	A^{n}_{m}  B^{n}_{m} = \alpha \nu^3
	$
and
	$
	E_m - E_n = h \nu
	$,
where the latter relates to
	Bohr's model.
In this form of Wien's law,
	$h$ is a scale factor
		relating time and energy
and signifies no quantization,
	as pointed out by Einstein.
The wavelength displacements
	providing the frequency $\nu$
result from
	the Doppler shifts of standing wave modes
caused by
	macroscopic motions of cavity walls
during 
	adiabatic compression or expansion
	\cite{Einstein1905a,Planck1914}.
The only assumptions of discreteness were therefore that
	molecules had discrete energy levels $E_n$,
so only
	the energy levels enter eq.~(\ref{e:einstein}),
and that
	their radiative interactions occurred as
		discrete events,
limiting
	the duration or length of the wave trains
		emitted in each transition.

The additional arguments of momentum
were necessary for
	completeness of the dynamical picture of interaction,
including
	the implication of induced Brownian motion
		from the radiative reactions.
Those ideas refined over
	his 1905 photoelectricity paper
	\cite{Einstein1905a},
which had concerned energy quanta
	but not momentum.
The arguments used the relation
	$p = E  c$
	from classical electromagnetics
	\footnote { 
	The momentum implication led Maxwell to accept
		Crooke's explanation of the radiometer
			that turned out wrong
		and was corrected by Reynold's theory
			in Maxwell's last paper
			(1879).
	The radiation pressure in
		both the Stefan-Boltmann and Wien's laws
	comes from 
		classical radiation momentum,
	and both Planck and Einstein
		depended on Wien's law.
	}, 
and the Lorentz property,
\begin{equation} \label{e:lorentz} 
	E' = E
		\sqrt{
			\frac{
				1 - v  c
			}{
				1 + v  c
			}
		}
\quad
\text{along with}
\quad
	\nu' = \nu
		\sqrt{
			\frac{
				1 - v  c
			}{
				1 + v  c
			}
		}
	.
\end{equation} 
which had been derived from
	a standpoint of energy conservation
		in his 1905 relativity paper
	\cite{Einstein1905c},
without involving
	any specific wave representation,
as opposed to
	its derivation in
	\cite{Popescu2007}.
The approach avoided
	convergence issues
historically encountered
	in Fourier theory
	\cite{Wheeler1987,Kleiner1989},
as well as
	problems of
		complexity and completeness of representation.

Einstein had been expressly concerned
	with ``the theory of light
		which operates with continuous spatial functions''
	in his photoelectricity paper
	\cite{Einstein1905a},
but his perception was
	classical and deterministic,
just as
	the action-angle formalism of quantum mechanics
to which he made key contributions
	\cite{Curtis2009}.
Eq.~(\ref{e:lorentz}) did not require or prove
	the irreducibility of quantization to classical laws,
despite its similarity to
	Planck quantization.
The dissociation from wave representations
	would have impeded
application of eq.~(\ref{e:lorentz}) to
	standing wave modes
		following Rayleigh and Planck,
however, just as
the strong notion of localization of radiant energy
	in deriving eq.~(\ref{e:lorentz})
went against
	the probabilistic quantum notions in
	\cite{EPR1935}.
In any case,
\emph{every} input in
	Einstein's derivation
was thus classical
\emph{except}
	the discreteness of molecular states and their transitions
	(eq.~\ref{e:einstein}),
so the derivation already implicated
	the discreteness of matter
		for quantization.

The classical derivation of eq.~(\ref{e:lorentz})
	needed for the present result
is adequately treated in
	\cite{Popescu2007},
but the further notion of quanta
	in \cite{Popescu2007}
as half-wavelength segments of travelling waves
	is not endorsed,
as it contradicts the uncertainty principle
	in single photon observations.
The only physical basis for that notion is
	the conservation of energy of
		any sequence of such segments,
\ie,
	of whole wave trains,
	under propagation
		and Lorentz transformations,
already assured classically
	without quantum significance.
Such half-wavelength segments of
	\emph{standing wave modes}
do correspond to quanta
	under second quantization
because
	harmonically related modes differ in geometry
		only by multiples of half-wavelengths,
	and in energy
		by whole quanta.
Along an internal dimension $L$
	in a cavity,
a standing wave mode of wavelength $\lambda$
	would have
	$
	N =
		2 L  \lambda
	=
		2 L \nu  c
	$
	such segments,
whence 
	$N$ is proportional to frequency $\nu$.
Planck's quantization rule
	$E = h \nu$
then promises
	$
	EN = 
		h \nu c  2 L \nu 
	=
		h c  2 L
	$,
independent of wavelength or mode,
	for segment energy,
and the quantum $h \nu$ is the energy of
	a whole mode of frequency $\nu$,
and not of
	an individual half-wavelength segment.


\Section{rayleigh}{Doppler-Lorentz correction to Rayleigh's theory} 

The further attempt in
	\cite{Prasad2006b}
to identify
	families of harmonically related
		standing wave modes
	with Planck oscillators,
		consistent with second quantization,
	was incomplete,
as it needed
	mode energies to correspond to
		integrals over phase increments in multiples of $\pi  2$,
	instead of
		the cavity dimension $L$.
The ordinary classical proportionality of
	standing wave energies to $L$
would leave the modes equiprobable under
	a Boltzmann probability distribution,
and thus reduce to
	Rayleigh's theory
		and its problem of divergence.
In hindsight,
eq.~(\ref{e:lorentz})
	also addresses that problem
since the premise of
	Doppler shifts due to wall vibrations
		as the mechanism of mode interactions
implies that
	a transition from a mode of frequency $N c  2 L$
		to frequency $(N + 1) \, c  2 L$
would also raise the energy by
	a ratio $(N + 1)N$,
taking up energy and momentum
	from the wall vibrations,
		and the \viceversa.
The energy of a mode of $N$ segments
	is then proportional to $N$,
		as required for Planck quantization
	per the reasoning above,
instead of being
	invariant of frequency,
as Rayleigh had assumed
	in ignorance of eq.~(\ref{e:lorentz}).
The corrected theory then yields
	the same cutoff at higher frequencies
		as Planck's,
as the higher frequencies would be less probable
	by the same probability factors
		because of the higher energies.

Since Planck's theory
	did not concern matter states,
and the standing wave modes,
	though discrete like the molecular states,
led to divergence
	in Rayleigh's theory,
Planck's inference of quantization derived from,
and thereby depended on,
	his hypothesis of oscillators
		with discrete energy levels.
This dependence is especially clear from
	the Bose derivation
	\cite{Bose1924},
which uses
	a dynamical phase space for
		the radiation energy and momentum
	discretized under
		Planck quantization
to eliminate
	Wien's law.
In his 1901 paper
	\cite{Planck1901},
Planck introduced
	a model of $N$ discrete resonators
that could bear only exact multiples of
	a small amount of energy $\epsilon$,
to compute
	the probability $W$ in Boltzmann's equation
		$S = k_B \log W$,
from
the number of ways
	the total energy $U_N = P \epsilon$
could be shared among
	the $N$ resonators.
This gave
\begin{equation} \label{e:oscentropy} 
	S
	=
		k
			( 1 + U  \epsilon )
			\log
			( 1 + U  \epsilon )
		-
			( U  \epsilon )
			\log
			( U  \epsilon )
\end{equation} 
for the per-resonator entropy
	in terms of the resonator energy $U$.
Separately,
Wien's law in the form
	$
	E \, d \lambda
	=
		\theta^5
		\gamma (\lambda \theta)
	\,
		d \lambda
	$
was shown to imply,
	via Kirchhoff-Clausius law
		for blackbody emissivity, 
the energy density 
\begin{equation} \label{e:wienthiesenplanck} 
	u
	=
		\frac{\nu^3}{c^3}
		f
		\left(
			\frac{\theta}{\nu}
		\right)
	,
\end{equation} 
where
	$\gamma$ and $f$ are functions
whose precise form
	is not significant in the result,
and
	$\theta$ is
		the thermodynamic temperature.
The energy $U$ of a resonator
	had been independently
related to
	the intensity of
		an applied linearly polarized oscillating field
	as
	$
	I = U \nu^2  c^2
	$,
which then led,
	via the relation
	$
	u = 8 \pi I  c
	$,
to
	$
	u
	=
		8 \pi \nu^2 I  c^3
	$,
so $U$ could be obtained independently of $c$,
	and related to temperature and frequency,
by combining with
	eq.~(\ref{e:wienthiesenplanck})
as
\begin{equation} \label{e:oscthermo} 
	U
	=
		\nu f \left(
			\frac{\theta}{\nu}
		\right)
\quad
\text{or, equivalently,}
\quad
	\theta
	=
		\nu f \left(
			\frac{U}{\nu}
		\right)
	.
\end{equation} 
The statistical definition of temperature
	$
	dS  dU = 1 / \theta
	$
then led to
	$
	S = f ( U  \nu )
	$
	with eqs.~(\ref{e:oscthermo}),
and to
	the proportionality
	$
	\epsilon \propto \nu
	$
on comparing with
	eq.~(\ref{e:oscentropy}).
Replacing
	$\epsilon$
with
	$h \nu$
in
	eq.~(\ref{e:oscentropy})
and evaluating
	the derivative $d S  d U$
then led to
\begin{equation} \label{e:planck} 
	U
	=
		\frac{h \nu}{
			e^{h \nu  k_B T} - 1
		}
\quad
\text{and, via the $u$-$I$-$U$ relations above,}
\quad
	u
	=
		\frac{8 \pi h \nu^3}{c^3}
		\frac{h \nu}{
			e^{h \nu  k_B T} - 1
		}
	.
\end{equation} 
Eqs.~(\ref{e:oscentropy}-\ref{e:planck})
	are directly reproduced from
	\cite{Planck1901}
but using symbol $I$ for the field intensity,
	instead of Planck's $K$,
and $k_B$ for Boltzmann's constant,
	to avoid confusion with wave numbers
		in the present treatment.

The resonator thermodynamic relations
	in eqs.~(\ref{e:oscthermo})
could not have implied
	discreteness of energy,
since
	eq.~(\ref{e:wienthiesenplanck}) comes from Wien's law
		whose considerations are strictly macroscopic,
and the only other consideration is
	the analogue field intensity $I$.
The discreteness and quantization
	in the result, eqs.~(\ref{e:planck}),
thus seem to originate from
	the oscillator model in eq.~(\ref{e:oscentropy}).
The proportionality implication is that
	the same factor $h$ must hold for all resonators
		in any cavity,
and also
	in all sets of cavities
		allowed to interact via radiation,
so $h$ must be
	a universal constant,
uniquely determined by
	the empirical energy density spectra,
		eqs.~(\ref{e:planck}),
	at different temperatures.

Yet, dimensionally,
	$h$ is a scale factor linking energy and frequency scales,
much as
	Boltzmann's constant $k_B$ relates
		energy and temperature
	and is also a universal constant,
and
Planck's model clearly followed
	Boltzmann's treatment of molecular energies
		in equal increments of $\epsilon$
	\cite{Boltzmann1877,Sharp2015},
which does not lead to
	a quantization of the molecular energies.
The result
	$\epsilon = h \nu$
does \emph{not} hold
	without eqs.~(\ref{e:oscthermo}) from Wien's law,
however,
so the cause of quantization in Planck's theory is again
	the atomic transitions producing the radiation
		in the observed spectra.
Bose's derivation showed that
	the nature of the frequency domain
makes
	$h$ irreducible to zero,
echoed in
	the quantum condition
	\cite{Born1926,Aitchison2008,Fedak2009},
but that aspect is implicit in
	the inherent anti-commutation of Poisson brackets
	\cite{Dirac1925}
	\cite[\S{21}]{Dirac}
and confers no more
	``experimental authority''
than say Hamilton already had for anticipating quantization
	(\cf \cite[\S{10-8}]{Goldstein}).

To verify this,
consider that in standing wave modes,
	discreteness is assured by definition,
and further,
any energy $\epsilon$ transferred from
	a mode of frequency $\nu$
to
	another of frequency $\nu' \ne \nu'$,
by simple reflection and Doppler shift
	from a moving wall or molecule,
or indirectly by
	its absorption by a molecule
		and subsequent re-emission
	with Doppler shift,
would be amplified to
	$
	\epsilon \nu'  \nu
	$
	by eq.~(\ref{e:lorentz}),
the difference
	$
	\epsilon (
		\nu' - \nu
		)
	$
coming from 
	the reflecting wall or molecule.
Equality of energy in
	the half-wavelength segments of a mode
is assured by
	wave propagation.
Assuming
	all frequency transitions
are governed by
	eq.~(\ref{e:lorentz}),
if the excitation energy $\epsilon$
	at frequency $\nu$
transits to
	a harmonically related mode of frequency $N \nu$,
by eq.~(\ref{e:lorentz}),
	the excitation at the harmonic would be
	$
	N \epsilon
	$,
	and the \viceversa.
If the transitions are due to
	classical interactions with molecules or the walls,
the harmonic excitation
	should occur with probability
	$
	e^{-N \epsilon  k_B T }
	$.
We should find
	the initial excitation
appears as
	energy $\epsilon$ at $\nu$
		with probability (or relative frequency) $e^{- \epsilon  k_B T}$,
	as $2 \epsilon$ at $2 \nu$
		with probability $e^{- 2 \epsilon  k_B T}$,
	and so on,
so the mean energy across
	harmonically related modes
becomes
\begin{equation} \label{e:swm} 
	\Exp{\epsilon}
	=
		\frac{
			\epsilon e^{- \epsilon  k_B T}
		+
			2 \epsilon e^{- 2 \epsilon  k_B T}
		+
			\dots
		}{
			e^{- \epsilon  k_B T}
		+
			e^{- 2 \epsilon  k_B T}
		+
			\dots
		}
	=
		\frac{
			\epsilon
		}{
			e^{\epsilon  k_B T} - 1
		}
	,
\end{equation} 
same as the expression for $U$ in
	Planck's result, eqs.~(\ref{e:planck}),
but with no reason to assume
	$\epsilon$ is quantized.
Eq.~(\ref{e:swm}) implies that
	harmonic families of standing wave modes
behave as
	Planck's harmonic oscillators
under equilibration by
	Doppler shifts.
The $-1$ in the denominator implies
	the same cutoff as Planck's law,
so Rayleigh's theory
	stands corrected.

Evaluation of the expectation
	over arbitrary sets of modes
would yield
	the same result,
since both
	molecular motions and wall vibrations
would equilibrate
	across all modes.
This too is easy to prove
	with the hindsight of past work.
The detailed balance between
	an arbitrary pair of modes 
of frequencies
	$\nu$
and
	$\gamma \nu$
then requires
	$
	\epsilon e^{- \epsilon  k_B T}
	=
		\gamma \epsilon e^{- \gamma \epsilon  k_B T }
	$
for a non-zero real ratio $\gamma$
	due to eq.~(\ref{e:lorentz}),
whence
	$
	\gamma
	=
		e^{ ( \gamma - 1 ) \epsilon  k_B T }
	$,
or equivalently
	$
	\gamma + 1
	=
		e^{ \gamma \epsilon  k_B T }
	$,
so that
	$
	(\gamma + 1)  \gamma
	=
		e^{ \epsilon  k_B T }
	$,
whereby the energy expectation is again
	$
	\Exp{ \epsilon }
	=
		\epsilon  [ e^{ \epsilon / k_B T } - 1 ]
	$,
along the lines of
	the Bose-Einstein derivation of Planck's law
	(\cf \cite[III-4-5]{Feynman}),
without depending even on
	the discreteness of the modes.
The constancy of $h$ in Planck's result,
	eqs.~(\ref{e:planck}),
thus indeed came from
	the atomic interactions leading to Wien's law.


\Section{theory}{Equilibrium and spectra of wave trains} 

Whereas Wien's law was
	a classical constraint
in obtaining
	the blackbody spectrum,
physicists have long followed the inverse position
	epitomized in the correspondence principle,
that classical physics is
	the long wavelength approximation of
		quantum mechanics.
The view reflects
	the traditional belief
that
	classical physics cannot possibly explain
		quantum behaviour,
whereas extrapolating
	the internal structure and properties of matter
		revealed by quantum mechanics
to explain
	macroscopic classical phenomena
look like
	a mere computational exercise.
While fundamental limits on such computations
	are also being discovered
	\cite{Cubitt2015},
our preceding result implies more than
	a mere equivalence of
		the power of classical physics,
since its reasoning involves
	inherently fewer assumptions.

The only new premise, that
	frequency transitions are governed by
		the Lorentz condition,
	eq.~(\ref{e:lorentz}),
had been already argued by Einstein in 1917
	as a condition for 
		dynamical consistency
	with the equilibrium of matter.
Only the further notion of
	equilibrating Doppler shifts
provided by wall vibrations
	is new,
but is more an intuition than a requirement,
	because eq.~(\ref{e:lorentz}) also holds with
		molecular absorptions and emissions
	in Einstein's reasoning.
The idea instead allows us to correct
	the defect of
		in Boltzmann's and Planck's theories
	of neglecting wall vibrations,
which would be available for equilibrating radiation
	even in a cavity with a perfect vacuum.

Also by Einstein's argument,
both molecular motions and wall vibrations
	should retain their classical equipartitions
requiring
	a distribution of $k_B T  2$ energy
		per degree of freedom,
but the standing wave modes are
	spectral components of the total radiation,
as well as
	macroscopic in geometry,
and for that reason,
	cannot exhibit mean energies of $k_B T  2$.
As wave trains emitted or absorbed by
	matter transitions
were pointed out as
	the real microscopic entities of radiation
		in thermal equilibrium with matter,
it might be thought that
	the wave trains must then possess $k_B T  2$ energies,
which too is impossible
	since the wave train energies
		must correspond initially to atomic transitions
	and do not change
		other than via atomic transitions,
	in Einstein's treatment.
Radiation is merely
	a means of exchange of energy and momentum
between
	the material constituents of
		the classical equipartition.
Like the stiffness of massless springs
	in a mechanical lattice,
a constraint on wave train energies
would affect
	the equilibration rate
but not
	the final equilibrium distribution.
All initial quantized energies and frequency spreads
	of the wave trains
are lost in
	the overall equilibrium,
so even wave trains,
	though particle-like
		in dynamical interactions,
are thermodynamically not at all
	particulate.

The derivation of
	the equilibrium spectrum for wave trains
below
illustrates
	another property of the Doppler effect
rarely discussed
	because of the traditional focus on individual frequencies,
		which correspond to Fourier components,
	instead of
		the composite wave trains
			they represent.
As derived in
	\cite{Einstein1905c}
and also explained in 
	\cite{Popescu2007},
the Lorentz transform implies
	\emph{quadratic} variation in
		electromagnetic field strengths
	with Doppler shifts,
but as the shifts themselves signify
	an inverse \emph{linear} variation in wavelengths
and thereby in the total \emph{duration} of
	any sequence of half-wavelength segments,
		meaning time dilations,
the energies delivered by the wave trains
	vary linearly with the shifts,
yielding
	eq.~(\ref{e:lorentz}).

The Doppler time dilations have been inconsequential
	for sinusoidal wave functions and Fourier components
because such functions are considered
	to extend to infinity.
They do affect
	timings in signal modulation,
as discovered recently the hard way
	\cite{Oberg2004,Oberg2005},
and admit propagating chirp modes
	that have been correspondingly overlooked
		in all of physics
	for three centuries
	\cite{Prasad2015b}.
The time dilations were not relevant
	in the reasoning of eq.~(\ref{e:swm})
as standing wave modes are Fourier components
	extending in time to $\pm \infty$.
The mode energies and probabilities
	were determined from
the varying number $N$ of half-wavelength segments
	under Doppler transitions
under
	a static cavity dimension $L$.
The lengths and
	amplitude distributions of wave trains
would be unchanged by reflections
	at strictly static walls
even as
	the wavefronts are spatially scattered,
but both would be affected by
	the Doppler shifts
		from vibrating walls.

Three further complications arise in considering
	wave trains.
Firstly,
only one standing wave mode is defined to exist
	at each location within a cavity
with given
	frequency,
	direction,
	polarization
	and angular momentum characteristics,
but any number of wave trains
	can share these properties,
analogous to
	photons in lasing.
As the spectral contributions of the wave trains
	merely superpose on one another
and wave trains
	do not exchange energy or momentum,
the spectral contributions of each wave train
	in equilibrium
must bear the same shape,
	and therefore follow eq.~(\ref{e:swm}).

Secondly,
eq.~(\ref{e:swm}) was derived for
	the thermalization of modes
describing
	the total radiation,
and a similar treatment seems needed for
	wave train spectra,
but eq.~(\ref{e:lorentz}) is only applicable to
	finite sequences of half-wavelength segments,
whereas
	a Fourier component of a wave train
		would have infinite segments.
A solution exists in
	the Dirichlet treatment for
		Fourier representations
	over a finite interval,
according to which
	the behaviour of components outside the interval
		should be immaterial.
The component energy
	beyond the wave train
		must be irrelevant because
the components must cancel out in phase
	beyond the wave train.
This notion is not diminished by
	the spatial dispersion of the wave fronts by
		the wall reflections,
since the energy delivered by a component
	at any given location
		is an integral over time.
With wall vibrations,
continuity of the wave trains
	is also not assured
because
the velocity of a reflecting wall will fluctuate
	over the duration and expanse of a wave train.
The \emph{temporal} volume of
	the original wave train
is unaffected except as
	the overall Doppler time dilation,
however,
so both
	time dilations
and
	the Lorentz property
are meaningful.

The third complication is that
	the probability factors of the ``oscillator expansion''
		represented in eq.~(\ref{e:swm}),
as well as
	in the corresponding detailed balance 
		of frequencies $\nu$ and $\gamma \nu$,
	concern different standing wave modes,
as in 
	Boltzmann's original reasoning for molecular energies
		in the kinetic theory,
but must now relate
	spectral components of the same wave train.
If a wave train occurred with probability $p$
	at overall energy $\epsilon$,
which got boosted to $\gamma \epsilon$
	by a Doppler shift of its spectrum,
the probability of observing it
	at the total energy $\gamma \epsilon$
		should be still $p$,
as Doppler shifts
	do not create or annihilate waves
		or wave spectral components.

However,
wave trains suffer fragmentation
	due to the wall vibrations,
and the fragments suffer
	differing, unrelated shifts at random,
so the multiplicative law
	$
	p(u)
	\equiv
		e^{- u  k_B T }
	=
		\exp(
			- \sum_i u_i  k_B T
		)
	\equiv
		\prod_i
			p(u_i)
	$,
holds, where $u_i$ denote
	fragment probabilities.
If a fragment bears
	energy $u$ at frequency $\nu$
		for a time interval $\tau$,
it can later appear
	with its original probability
	$
	e^{- u  k_B T }
	$
at frequency
	$\nu' = \gamma \nu$,
under
	a Doppler shift factor
		$\gamma \ne 1$
only with energy
	$
	u'
	=
		u \omega'  \omega
	\equiv
		\gamma u
	$
	due to the Lorentz increase
and only for
	the shorter period
	$
	\tau'
	=
		\tau
		\;
		\omega  \omega'
	\equiv
		\tau  \gamma
	$.
To be observed at $\nu'$ for
	the full original interval $\tau$,
the fragment must
	\emph{repeatedly appear} at $\nu'$
		$\gamma$ times in succession,
which has
	the product probability
	$
	e^{- \gamma u  k_B T }
	$.
This is indeed how standing wave modes at
	frequencies $\nu$ and $\gamma \nu$
		are related,
as the $\gamma \nu$ mode corresponds to
	$\gamma$ repetitions of
		the $\nu$ mode \emph{in phase}
	over their common length $L$
		given by the cavity geometry.
Setting
	$\gamma = 2, 3 \dots$
reproduces
	the terms of the ``oscillator expansion''
		in eq.~(\ref{e:swm}),
	and hence the same result.


\section{Conclusion}

Despite reproducing
	Planck's result
and
reinforcing
	his own heuristic of radiation quanta
		in photoelectricity theory
	with momentum arguments,
Einstein's 1917 derivation was
	rigorous reasoning of classical dynamics,
and its prediction of
	stimulated emission,
for instance,
	would have been unconvincing otherwise.
It traced quantization to
	the discreteness of atomic states
and
	constancy of
		their transition energies,
which means conversely that
the equilibrium should exhibit
	the Planck form without quantization
if matter state transitions
	were excluded.
This converse implication has been proved above for
	the equilibration
both of
	standing wave modes
and of
	arbitrary sets of wave trains.

That
Einstein did not emphasize
	this fundamental implication of
		matter interactions
	in his derivation
		of Planck's law,
and
	of the Lorentz property
		that he himself had discovered,
and instead endorsed
	Bose's derivation
and formulated
	Bose-Einstein statistics,
is consistent wtih
	his focus on
		the internal structure of matter.
It is at most
	a error of omission
in not examining
	the converse implication,
and an informal concern
	against the present result.

More particularly,
the physicists' ignorance of
	the classical Lorentz property,
and not an inability of
	classical physics,
has been shown the real reason for
	Rayleigh's failure in obtaining
		the correct law.
Likewise,
the inattention to matter interactions,
	which enter only via Wien's law 
		and were addressed only later by Einstein,
has been shown to have left
	Planck unable to distinguish
the origin of quantization from 
	a computational artefact in estimating entropy
		also used by Boltzmann.
An even older defect,
neglect of
	thermal vibrations of the bounding walls
in direct contact
	with a gas or radiation
		stipulated to be in equilibrium,
has been also identified
	and corrected with the converse insight.

---------------------------------------------------------------------
\raggedright

\end{document}